# Non-trivial band topology and orbital-selective electronic nematicity in a new titanium-based kagome superconductor

Yong Hu[1,#,*], Congcong Le[2,#], Zhen Zhao[3,4], Junzhang Ma[5,6], Nicholas C. Plumb[1], Milan Radovic[1], Andreas P. Schnyder[7], Xianxin Wu[8,*], Hui Chen[3,4], Xiaoli Dong[3,4], Jiangping Hu[3,4], Haitao Yang[3,4], Hong-Jun Gao[3,4,*], and Ming Shi[1,*]

[1] *Photon Science Division, Paul Scherrer Institut, CH-5232 Villigen PSI, Switzerland*
[2] *RIKEN Interdisciplinary Theoretical and Mathematical Sciences (iTHEMS), Wako, Saitama 351-0198, Japan*
[3] *Beijing National Center for Condensed Matter Physics and Institute of Physics, Chinese Academy of Sciences, Beijing 100190, PR China*
[4] *School of Physical Sciences, University of Chinese Academy of Sciences, Beijing 100190, PR China*
[5] *Department of Physics, City University of Hong Kong, Kowloon, Hong Kong, China*
[6] *City University of Hong Kong Shenzhen Research Institute, Shenzhen, China*
[7] *Max-Planck-Institut für Festkörperforschung, Heisenbergstrasse 1, D-70569 Stuttgart, Germany*
[8] *CAS Key Laboratory of Theoretical Physics, Institute of Theoretical Physics, Chinese Academy of Sciences, Beijing 100190, China*

[#] These authors contributed equally to this work.

* To whom correspondence should be addressed:

Y.H. (yonghphysics@gmail.com);
X.X.W. (xxwu@itp.ac.cn);
H.-J.G. (hjgao@iphy.ac.cn);
M.S. (ming.shi@psi.ch).

**Electronic nematicity that spontaneously breaks rotational symmetry has been shown as a generic phenomenon in correlated quantum systems including high-temperature superconductors [1-9] and the $A$V$_3$Sb$_5$ ($A$ = K, Rb, Cs) family with a kagome network [10]. Identifying the driving force has been a central challenge for understanding nematicity. In iron-based superconductors, the problem is complicated because the spin, orbital and lattice degrees of freedom are intimately coupled [11]. In vanadium-based kagome superconductors $A$V$_3$Sb$_5$, the electronic nematicity exhibits an intriguing entanglement with the charge density wave order (CDW), making understanding its origin difficult. Recently, a new family of titanium-based kagome superconductors $A$Ti$_3$Bi$_5$ has been synthesized [12]. In sharp contrast to its vanadium-based counterpart, the electronic nematicity occurs in the absence of CDW [13,14]. $A$Ti$_3$Bi$_5$ provides a new window to explore the mechanism of electronic nematicity and its interplay with the orbital degree of freedom. Here, we combine polarization-dependent angle-resolved photoemission spectroscopy with density functional theory to directly reveal the band topology and orbital characters of the multi-orbital RbTi$_3$Bi$_5$. The promising coexistence of flat bands, type-II Dirac nodal line and nontrivial $\mathbb{Z}_2$ topological states is identified in RbTi$_3$Bi$_5$. Remarkably, our study clearly unveils the orbital character change along the $\overline{\Gamma}$-$\overline{M}$ and $\overline{\Gamma}$-$\overline{K}$ directions, implying a strong intrinsic inter-orbital coupling in the Ti-based kagome metals, reminiscent of iron-based superconductors. Furthermore, doping-dependent measurements directly uncover the orbital-selective features in the kagome bands, which can be well explained by the $d$-$p$ hybridization. The suggested $d$-$p$ hybridization, in collaboration with the inter-orbital coupling, could account for the electronic nematicity in $A$Ti$_3$Bi$_5$. As such, our finding suggests a promising mechanism for understanding electronic nematicity that presents in correlated systems and most probably competes with superconductivity.**

Electronic nematicity and its fluctuations, as a ubiquitous normal state property, are arguably linked to emergent superconductivity [1-9,15]. However, understanding its driving force has been a central challenge for correlated quantum systems. The kagome lattice, a corner-sharing triangle network, has emerged as one of the most fundamental systems for studying exotic correlated and topological quantum states. Due to its frustrated lattice geometry and unique correlation effects embedded in flat bands and van Hove singularities [16-23], a wide range of electronic instabilities and nontrivial topologies have been observed, including quantum spin liquid [24-27], Dirac/Weyl semimetals [29-31], charge density wave (CDW) orders [16-18], and unconventional superconductivity [16-18,32]. Within this realm, vanadium-based superconductors $A$V$_3$Sb$_5$ ($A$=K, Rb, Cs) [33,34] have attracted much recent attention because they exhibit intriguing similarities to correlated electronic phenomena observed in high-temperature superconductors, such as a pair density wave [35], time-reversal symmetry broken CDW [36,37], and electronic nematicity [10]. However, the nematic order in $A$V$_3$Sb$_5$ is entangled with the CDW, making understanding its origin challenging.

Very recently, a new family of Ti-based kagome metals $A$Ti$_3$Bi$_5$, which is isostructural with $A$V$_3$Sb$_5$, has been synthesized [12]. Despite a distinct 3$d$ electron configuration in Ti atoms, superconductivity occurs with an onsite transition temperature $T_c$~4.8 $K$. The strong spin-orbit coupling (SOC) from Bi atoms can generate intriguing nontrivial topological phenomena. In stark contrast to $A$V$_3$Sb$_5$, transport measurements on $A$Ti$_3$Bi$_5$ show no evidence of a CDW state [12,38]. Interestingly, an electronic nematicity with the rotational symmetry breaking is discovered in the absence of the concomitant translation symmetry breaking [13,14], similar to iron-based high-temperature superconductors. Therefore, $A$Ti$_3$Bi$_5$ is a tantalizing system for understanding the mechanism behind electronic nematicity and its interplay with intertwined correlated quantum phenomena such as superconductivity.

RbTi$_3$Bi$_5$ crystallizes in a layered hexagonal lattice consisting of alternately stacked Ti-Bi sheets and Rb layers [Fig. 1a(i)]. It shares the same crystal structure as $A$V$_3$Sb$_5$, but with a kagome net of Ti atoms replacing the V. The corresponding bulk Brillouin zone (BZ) and the projected two-dimensional (2D) BZ on the (001) surface are illustrated in Fig. 1b, with high-symmetry points indicated. The band structures of RbTi$_3$Bi$_5$ from DFT calculations without/with SOC are displayed in Fig. 1c (also Figs. 2b and 2c). The characteristic band feature of the kagome lattice, e.g., the flat band emerges around a binding energy ($E_B$) of 0.3 $eV$ (as highlighted by the gray-colored region) and Dirac point (DP) occurs above the Fermi level ($E_F$) at the $\bar{K}$ point (indicated by the red arrow and circle). A detailed examination of the band structure of the Ti-based kagome metals along the out-of-plane momentum reveals a type-II Dirac nodal line (DNL, as marked by the green circle in Fig. 1c and Fig. 2b, for details see Fig. S1). The type-II DNL is protected by a mirror symmetry when the SOC is ignored. With further inclusion of SOC, a negligible gap would open at the type-II DNL, distinct from the pronounced gaps that open at the DP at the $\bar{K}$ (Fig. 1c) and trivial crossing points of the band structure in the momentum space (representatively shown by the black arrows in Figs. 2b and 2c). To gain insight into the electronic nematicity and search for the non-trivial band topology, we employ polarization-dependent angle-resolved photoemission spectroscopy (ARPES) to systematically study the electronic structures of single-crystal RbTi$_3$Bi$_5$.

In Fig. 2d(i), we first present the measured Fermi surface (FS) sheets, which consist of one circle-like and two hexagonal-like electron pockets near the zone center ($\bar{\Gamma}$ point), one rhombic-like hole pocket at the zone boundary ($\bar{M}$) and one triangle-like electron pocket near the zone corner ($\bar{K}$), which agrees well with the theoretical calculations (Fig. 2a). The dispersive nature of the bands contributing to the FS is revealed along two representative high-symmetry paths, the $\bar{\Gamma}$-$\bar{M}$ (Fig. 2e) and $\bar{\Gamma}$-$\bar{K}$ (Fig. 2f) directions. (The four bands crossing the $E_F$ are denoted as α, β, γ and δ hereafter.) Due to the multi-orbital nature of Ti-$d$ orbitals, ARPES measurements of the band structure are strongly sensitive to the photon polarization. In principle, according to the selection rules in photoemission [39], the bands can be selectively detected depending on their symmetry with respect to the mirror plane formed by the photon beam, sample, and spectrometer (Fig. 2g). By exploiting these selection rules, the bands' characters can be determinized experimentally (see Fig. S2 for details of the matrix element analysis and Supplementary Table I for the summary of the detectable Ti-$d$ orbitals under our ARPES geometry). For instance, the β and γ bands are selectively detected by the linear vertical (LV) polarization [Fig. 2e(iii)] along the $\overline{\Gamma}$-$\overline{M}$ direction, indicating an odd symmetry (i.e., $d_{yz}$ and $d_{xy}$ orbitals) with respect to the mirror plane; meanwhile, along the $\overline{\Gamma}$-$\overline{K}$ direction, the β and γ bands are favored by LV polarization [Fig. 2f(iii)] and linear horizontal (LH) polarization [Fig. 2f(ii)], respectively. If the β and γ bands retain their orbital characters along the $\overline{\Gamma}$-$\overline{M}$ and $\overline{\Gamma}$-$\overline{K}$ directions, they would in principle be assigned to the $d_{xy}$ and $d_{yz}$ orbitals, respectively (Fig. S2). However, these apparently contradict the orbital characters from the theoretical calculation [13] (see also Fig. S2). This motivates us to consider the inter-orbital coupling between the non-degenerate orbitals ($d_{yz}/d_{xz}$, and $d_{xy}/d_{x^2-y^2}$) in the kagome lattice. In Fig. 2h, we show the calculated orbital-resolved band dispersion originating from sublattice A [Fig. 1a(ii)], invariant under mirror reflection $M_{xz}$ and $M_{yz}$, where the β (γ) bands along the $\overline{\Gamma}$-$\overline{M}$ and $\overline{\Gamma}$-$\overline{K}$ directions are mainly comprised of $d_{yz}$ ($d_{xy}$) and $d_{xz}$ ($d_{x^2-y^2}$) orbitals, respectively. This is consistent with the results from the polarization-dependent measurements [Figs. 2e(ii)-(iii), Figs. 2f(ii)-(iii) and Fig. S2], implying a strong intrinsic inter-orbital coupling in the Ti-based kagome metals.

The experimental band dispersions (Figs. 2e and 2f) also show good overall agreement with the DFT calculations (Fig. 1c, Figs. 2b and 2c). As shown in Figs. 2e (i) and (iii), as $E_B$ increases, the β and γ bands separated at $E_F$ ([Fig. 2d (i), Figs. 2e (I) and (iii)) evolve to intersect near $E_B$ = 0.25 $eV$ [Fig. 2d(ii), Figs. 2e (I) and (iii)], forming the predicted type-II DP (Fig. 2b, see Fig. S1 for details). As expected from the destructive interference of hopping in the frustrated kagome lattice (Fig. 2i) and the DFT calculations in RbTi$_3$Bi$_5$ (Fig. 1c, Figs. 2b and 2c), a strikingly nondispersive feature near the $E_F$ around $E_B$ =0.25 $eV$ is revealed [Figs. 2e(i) and 2f(i)]. Accordingly, as shown in the constant energy contour (CEC) at $E_B$ =$E_1$ in Fig. 2d(ii), the spectral weight of this feature is uniformly distributed across almost the entire 2D momentum space [Fig. 1c and Fig. 2is (ii)]. The flat feature is more evident under LV polarized light [Figs. 2e(iii), 2f(iii) and 2j] and further evidenced by the non-dispersive peak in the energy distribution curves (EDCs) (indicated by the red box in Figs. 2k and 2l).

Besides the type-II DNL and flat bands, the high resolution in the ARPES measurement allows us to reveal some fine structures in the band dispersions. As shown in Figs. 2f(i) and (ii), a double-band splitting is observed in the γ and ε bands along the $\overline{\Gamma}$-$\overline{K}$ path and is accentuated by the second derivative plot [marked by the arrows in Fig. 2f(ii)]. Furthermore, the effect of SOC on the band dispersions, i.e., the SOC associated gaps, are directly seen [as indicated by the black arrows in Figs. 2e(i) and 2f(i)]. Apparently, the DFT calculated bands with SOC (Figs. 2b and 2c) can better capture the experimental band structure in RbTi$_3$Bi$_5$ (Figs. 2e and 2f). The prominent SOC promotes the existence of the topologically nontrivial Dirac surface states (TDSSs) predicted by the *ab initio* calculations [12].

TDSSs originating from a $\mathbb{Z}_2$ bulk topology are indeed observed in RbTi$_3$Bi$_5$ (Fig. 3). A series of experimental CECs in Fig. 3a, measured from +0.18 *eV* to −0.12 *eV* with respect to $E_{DP}$, clearly show the evolution of the TDSSs in the energy space around the $\overline{\Gamma}$ point. To closely visualize the momentum-space structure of the Dirac bands, we show the band dispersions along two different high-symmetry directions, $\overline{\Gamma}$ - $\overline{M}$ (Fig. 3b) and $\overline{\Gamma}$ - $\overline{K}$ (Fig. 3c). Photon energy-dependent measurements reveal that Rashba-like bands [Fig. 3b(ii)] around the $\overline{\Gamma}$ point do not disperse with respect to photon energy (and thus $k_z$), in contrast to the bulk states (for details, see Fig. S3), confirming their 2D nature. To further explore the topological nature of the Rashba-like feature, we show the calculated bulk states projected onto the (001) surface (Fig. 3d) together with the theoretical surface spectra (Fig. 3e). The TDSSs derived from bulk nontrivial topology are theoretically identified around the $\overline{\Gamma}$ point [Fig. 3e(ii)]. Since the energy position and the shape of the surface states are highly sensitive to the details of the surface environment, the energy position of the DP is deeper in experiment [Fig. 3b(ii)] than in the calculations [Fig. 3e(ii)].

After characterizing the orbital characters and identifying the rich non-trivial band topology, we now demonstrate direct manipulation of the electronic states via *in situ* surface potassium (K) deposition [Fig. 4a(i)]. As shown in Fig. 4a(ii), the successful introduction of K atoms doped on the sample surface is confirmed by measuring the K 3*p* core level (red curve) which is absent on the pristine surface (black curve). To show the overall doping evolution of the band structure, we display the doping-dependent ARPES spectra in volume plots (Fig. 4b). Remarkably, upon doping, the top of the δ band is tuned from well above to below $E_F$, as shown in the high-symmetry band dispersions along the $\overline{\Gamma}$-$\overline{M}$ (Fig. 4c) and $\overline{\Gamma}$-$\overline{K}$ (Figs. 4d and 4e). This indicates that a Lifshitz transition of the FS driven by the δ band is realized with doping (Fig. 4b, see Fig. S4 for details). A comparative examination of the EDCs, taken at the $\overline{M}$ point, $\overline{K}$ point and the flat band, indicates the band shift does not shift in a simple, rigid fashion (Fig. 4f); rather, the shift is strongly orbital-dependent. Specifically, after doping, the hole-like ζ$_1$ and ζ$_2$ bands around the $\overline{M}$ point shift by about 60 *meV* and 90 *meV* [EDC#1, Fig. 4f(i)], respectively; the hole-like θ band near the $\overline{K}$ point moves down by 320 *meV* [(EDC#2, Fig. 4f(ii)]; and the flat band (η) along the $\overline{\Gamma}$-$\overline{K}$ path only drops by about 40 *meV* [EDC#3, Fig. 4f(iii)].

We next turn to discuss the implication of the observed orbital-selective doping effect on the kagome bands. As the kagome layers are embedded between two Bi layers [Fig. 1a(i)], upon surface deposition, the electron doping in the kagome layers is mediated by the coupling between Bi-*p* and Ti-*d* orbitals. The theoretically calculated orbital-resolved band structure distinguishing Ti-*d* and Bi-*p* orbitals, as shown in Fig. 4g, clearly reveals a strong *d-p* hybridization around the $E_F$ in RbTi$_3$Bi$_5$ (see Fig. S5 for details). The mirror-even (odd) $d_{xy}/d_{x^2-y^2}$ ($d_{xz}/d_{yz}$) orbitals under the mirror reflection $M_{xy}$ in the Ti-kagome plane couple with the $p_{x/y}$ bonding (anti-bonding) orbitals of honeycomb Bi atoms above and below. Interestingly, the *d-p* coupling along the $\overline{\Gamma}$-$\overline{K}$ path is stronger than along the $\overline{\Gamma}$-$\overline{M}$. The bands (θ and electron-like γ) around the $\overline{K}$ point and the hole-like δ band around the $\overline{M}$ point have the strongest *d-p* coupling, echoing the pronounced energy shifts observed on these bands upon surface K deposition [Figs. 4c-4e and 4f (ii)]. This *d-p* scenario can also account for the revealed double band splittings (in the γ and ε bands) along the $\overline{\Gamma}$-$\overline{K}$ path [Figs. 2f (i) and (ii)], one of which could be assigned to the surface band and the other to bulk states. As the exposed honeycomb Bi2 layers [Fig. 1a(i)] experience a surface potential, the surface bands with strong *d-p* coupling can be split from their corresponding bulk bands, resulting in a double band splitting.

Reminiscent of our finding, prominent nematic features in the STM measurement are observed from the intra-band scattering between the electron pocket (i.e., the γ band with $d_{xz}/d_{yz}$ orbitals) around the $\overline{K}$ point (wavevector $q_3$ in Fig. 4h) and between the $d_{xy}/d_{x^2-y^2}$ electron pocket around the $\overline{\Gamma}$ point (wavevector $q_4$ in Fig. 4h) [13,14], and the corresponding quasi-particle interference (QPI) pattern along $\overline{\Gamma}$-$\overline{K}$ appears to have stronger $C_6$-symmetry-breaking signatures. The electronic states connected by the $q_3$ and $q_4$ wavevectors show pronounced *d-p* coupling (Fig. 4g), suggesting the important role of *d-p* hybridization in promoting nematicity. Moreover, the QPI at $q_3$ ($q_4$) involves both intra-orbital and inter-orbital $d_{xz}/d_{yz}$ ($d_{xy}/d_{x^2-y^2}$) scattering. In the kagome lattice with three sublattices, the site symmetry is $D_{2h}$ and thus all five orbitals on each site are non-degenerate. In hexagonal systems, nematicity has a unique three-state Potts character [40,41], distinct from the Ising-like features in tetragonal systems. Accordingly, the simplest nematic order to break the $C_6$ rotational symmetry is an onsite sublattice potential, but it induces a uniform symmetry-breaking feature in the momentum space, which is inconsistent with the STM measurements [13,14]. Given the observed inter-orbital coupling (Figs. 2a, 2e-2h), an alternative scenario is intra- or inter-orbital bond order, where a stronger bonding (indicated by the thick line in Figs. 4i-k) between certain sites breaks the six-fold rotational symmetry but preserves the two-fold rotational symmetry, as illustrated in Figs. 4i-k (take the $d_{xz}/d_{yz}$ orbitals as an example). It is expected to display noticeable momentum-dependent nematic features. The effective hopping between *d* orbitals through Bi-$p_{x/y}$ orbitals can become nematic once the degeneracy of the $p_{x/y}$ orbitals is lifted. Importantly, a strong *d-p* hybridization can make the Ti 3*d* orbitals more extended, and thereby enhance the non-local Coulomb interaction, which can promote the nematic bond order [17,42].

Combined with DFT calculations, our ARPES results identify the promising coexistence of flat bands and nontrivial band topology (i.e., type-II DNL and nontrivial $\mathbb{Z}_2$ topology) in the Ti-based kagome superconductors RbTi$_3$Bi$_5$. These observations call for future investigations into their topological contributions to transport properties and related potential applications. Remarkably, by working with polarization-dependent ARPES, our study reveals the intricate orbital character change along the $\overline{\Gamma}$-$\overline{M}$ and $\overline{\Gamma}$-$\overline{K}$ directions (Fig. 2a and 2e), implying a strong inter-orbital coupling in the Ti-based kagome lattice in contrast to $A$V$_3$Sb$_5$ [23] (Fig. S6). This is reminiscent of iron-based superconductors, where $d_{xz/yz}$ orbitals are degenerate in tetragonal systems. The two sets of $d_{xz/yz}$ and $d_{xy/x^2-y^2}$ orbitals in RbTi$_3$Bi$_5$, due to the $D_{2h}$ site symmetry group, are nondegenerate, making the Ti-based kagome metals unique and distinct from iron-based superconductors. Furthermore, doping dependent measurements directly uncover the orbital-selective characters in the multi-orbital kagome system. The revealed *d-p* hybridization, in collaboration with the inter-orbital coupling, could provide a qualitatively account for the electronic nematicity in $A$Ti$_3$Bi$_5$. Thus, our finding would provide valuable insights into the understanding of nematic orders that present in correlated systems and most probably compete with superconductivity.

*Note:* As we were finalizing our manuscript, we became aware of similar experiments posted as preprints [43-45].

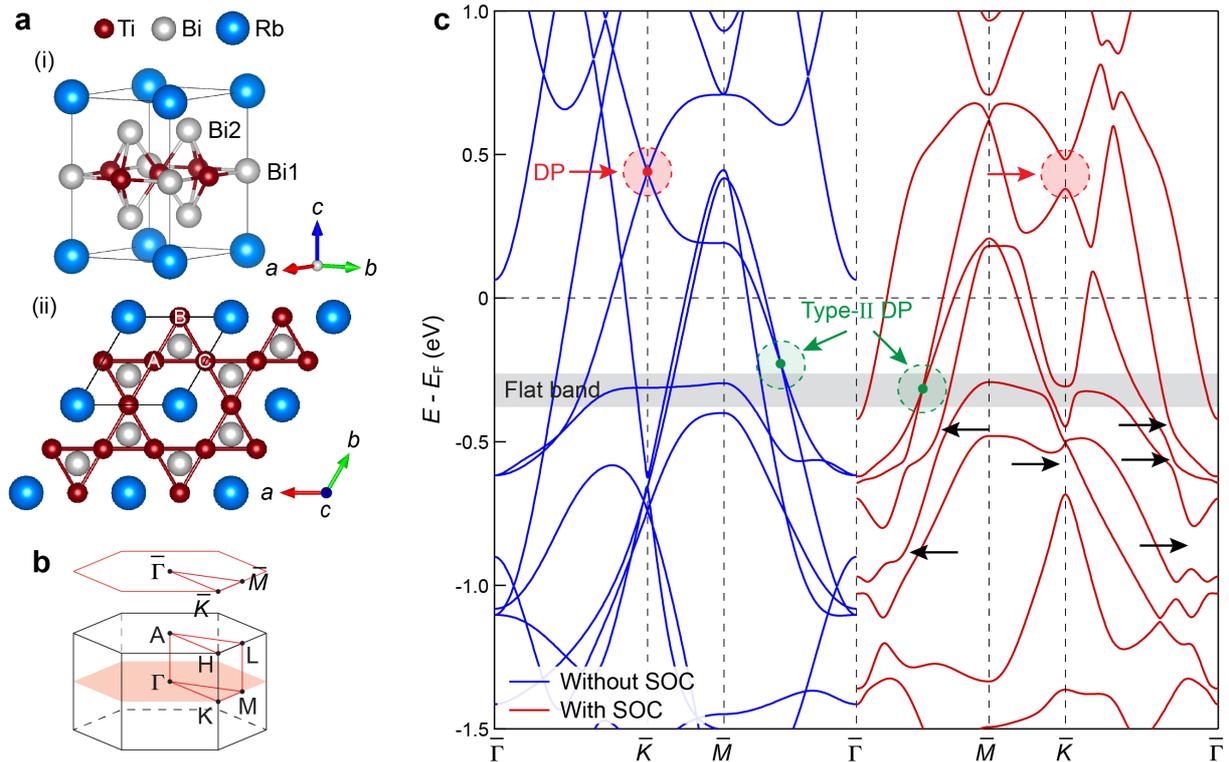

**Fig. 1 | Crystal structure and calculated band structure of RbTi$_3$Bi$_5$. a** The unit cell of RbTi$_3$Bi$_5$ with two types of Bi atoms indicated (i) and top view showing the kagome plane (ii). The white letters (A,B,C) in (ii) indicate the three sublattices in the kagome lattice. **b** Bulk Brillouin zone (BZ) of RbTi$_3$Bi$_5$ and the projection of the (001) surface BZ. **c** Density functional theory (DFT) calculated electronic structure of RbTi$_3$Bi$_5$ without spin-orbital coupling (SOC) (blue) and with SOC (red). The arrows in (c) indicated the Dirac point (DP, shown in red), Type-II DP (green), and SOC gap (black). The gray shading highlights the flat bands.

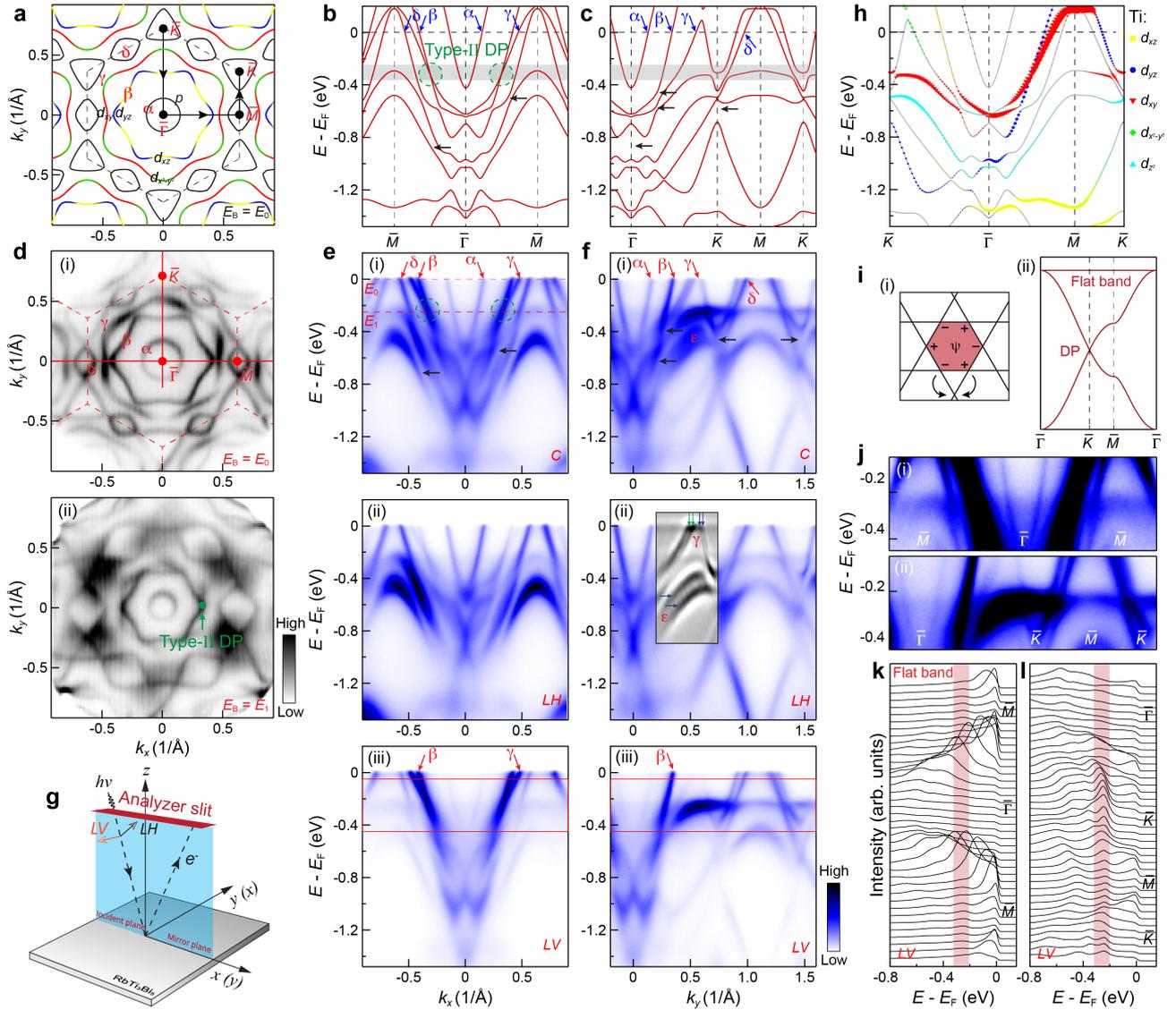

**Fig. 2 | Polarization-dependent measurements of the band structure and kagome flat band.** **a** Calculated Fermi surface (FS). Dashed hexagon represents the BZ. **b,c** DFT band structure along the $\bar{\Gamma}$-$\bar{M}$ (b) and $\bar{\Gamma}$-$\bar{K}$ (c) directions. **d** Constant energy contours (CECs) at the Fermi level ($E_F$) (i) and flat band (ii). The green arrow in (ii) marks the type-II DP. **e** ARPES spectra along the $\bar{\Gamma}$-$\bar{M}$ direction, probed with circular (C)(i), linear horizontally (LH)(ii), and linear vertically (LV)(ii) polarized light. The red dashed line in (i) marks the energy position of the CECs in (d). **f** Same as (e), but measured along the $\bar{\Gamma}$-$\bar{K}$ direction. The inset in (ii) highlights the band splitting in the γ and ε bands. **g** Experimental geometry of the polarization-dependent ARPES measurements. **h** Orbital-resolved DFT band dispersion originating from sublattice A [as indicated in Fig. 1a(ii)], along the $\bar{K}$-$\bar{\Gamma}$-$\bar{M}$-$\bar{K}$ path [as indicated by the black arrow in (a)]. The black arrow in (b-f) marks the SOC gap. **i** The destructive interference of hopping in the frustrated kagome lattice (i), as well as the expected flat band and DP (ii). **j** Zoomed-in plots of the flat band shown in [e(iii)] (i) and [f(iii)] (ii), as highlighted by the red box in (e) and (f). **k** Energy distribution curves (EDCs) of [e(iii)]. **k** EDCs of [f(iii)].

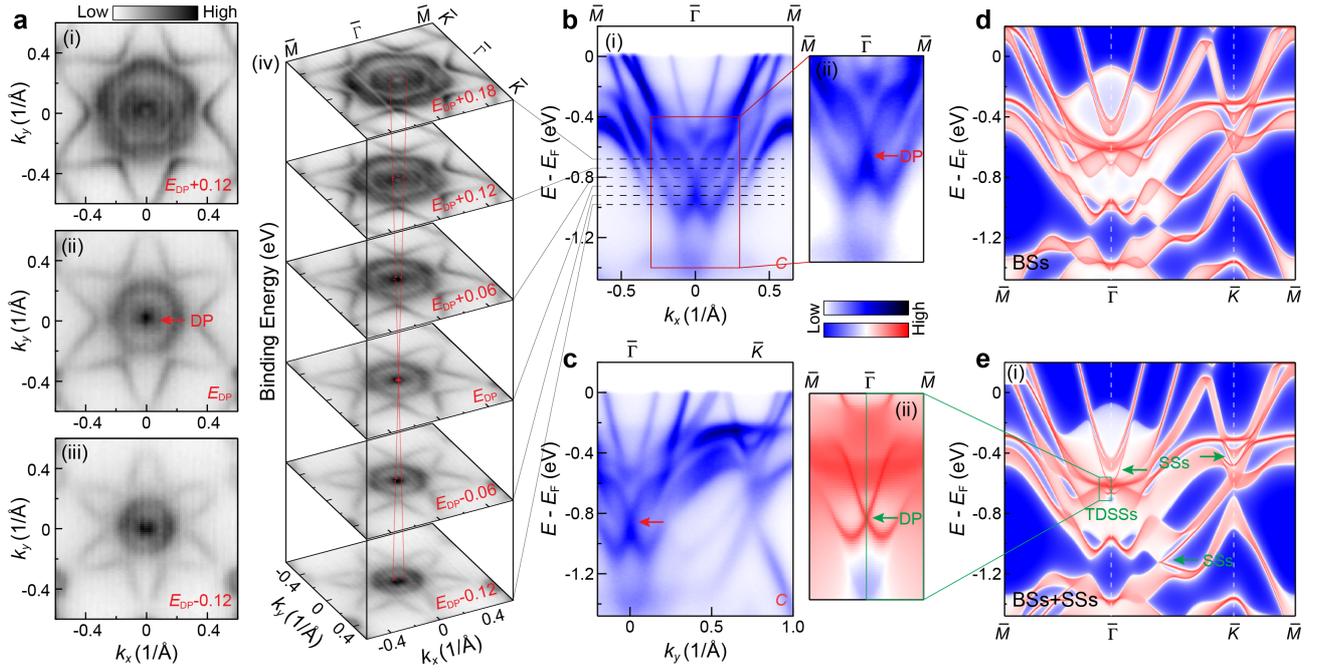

**Fig. 3 | $\mathbb{Z}_2$ Topological surface states in RbTi$_3$Bi$_5$. a** CECs at +0.12 eV (i), 0 eV (ii), −0.12 eV (iii) with respect to the $E_{DP}$, as well as the stacking plots (iv). **b** Experimental band dispersion along the $\overline{\Gamma}$-$\overline{M}$ direction (i) and enlarged plot of the dispersion near the DP (ii). **c** Same as (b), but along the $\overline{\Gamma}$-$\overline{K}$ direction. **d,e** The (001) surface Green's function projection of pure bulk states (BSs) (d) and the theoretical surface spectra [BSs and surface states (SSs)] (e). Enlarged plot of the calculated dispersion [e(i)] showing the topologically nontrivial Dirac surface states (TDSSs) is shown in [e(ii)]. Besides the TDSSs, the calculations also identify SSs along the $\overline{\Gamma}$-$\overline{K}$ path [as highlighted by the green arrows in e(i)], however, these states may overlap with BSs.

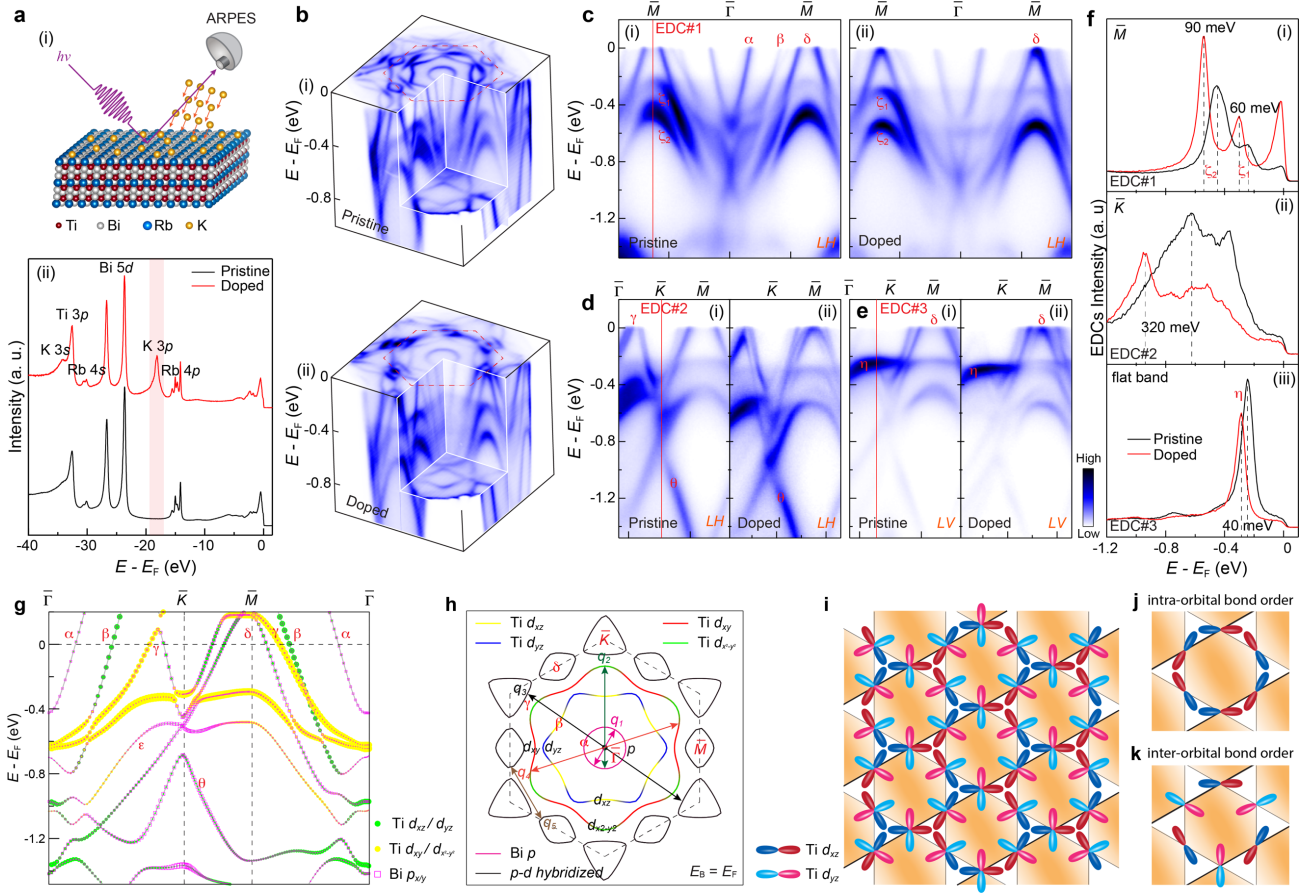

**Fig. 4 | Orbital-selective doping effect and *d-p* hybridization in RbTi₃Bi₅.** **a** Sketch of the *in-situ* potassium (K) deposition (i) and doping dependence of core-level photoemission spectrum showing the characteristic Rb 4*s*, Bi 5*d* and Ti 3*p* peaks (ii). Upon K deposition, the K 3*p* peak emerges (red curve), which is absent on the pristine surface (black curve). **b** Three-dimensional intensity plot of the electronic structure measured on the pristine (i) and K doped (ii) surfaces. **c** Doping evolution of the band dispersion along the $\overline{\Gamma}$-$\overline{M}$ direction measured on the pristine (i) and K doped surfaces (ii), probed with LH polarized light. **d,e** Same as (c), but measured along the $\overline{\Gamma}$-$\overline{K}$ direction with LH (d) and LV (e) polarizations. **f** Doping dependent EDCs taken around $\overline{M}$ point (i), $\overline{K}$ point (ii) and flat band (iii). The momentum location of the EDCs (#1-#3) is marked by the red line in (c-e). **g** Ti-*d* and Bi-*p* orbital-resolved DFT band dispersion. **h** Calculated FS showing four pockets contributed by different orbital bands. The scattering vectors ($q_1$-$q_5$) are the wavevectors of the quasi-particle interference patterns in scanning tunneling microscopy/spectroscopy measurements [13]. **i-k** The inter-orbital and intra-orbital couplings in the Ti-based kagome lattice (i), the intra-orbital band order (j) and inter-orbital band order (k). The yellow background in (i-k) represents the rotational symmetry breaking.